\documentclass[12pt,preprint]{aastex}
\usepackage{epsfig}

\begin{document}

\title{Observational Properties of Proto-planetary Disk Gaps}

\author{
Peggy Varni\`ere\altaffilmark{1}, J.~E.~Bjorkman\altaffilmark{2},
Adam Frank\altaffilmark{1}, Alice C.~Quillen\altaffilmark{1},
A.~C.~Carciofi\altaffilmark{2}, Barbara
A.~Whitney\altaffilmark{3}, and Kenneth Wood\altaffilmark{4}}

\altaffiltext{1}{Department of Physics \& Astronomy, Rochester University,
Rochester NY 14627-0171; pvarni@pas.rochester.edu; aquillen@pas.rochester.edu; afrank@pas.rochester.edu.}
\altaffiltext{2}{Ritter Observatory, MS 113, Department of Physics and Astronomy, University of Toledo, Toledo, OH 43606-3390;
jon@physics.utoledo.edu; acarcio@physics.utoledo.edu.}
\altaffiltext{3}{Space Science Institute, 3100 Marine Street, Suite A353,
Boulder, CO 80303; bwhitney@colorado.edu.}
\altaffiltext{4}{School of Physics \& Astronomy, University of St Andrews,
North Haugh, St Andrews, Fife, KY16 9AD, Scotland the Brave;
kw25@st-andrews.ac.uk.}

\begin{abstract}
We study the effects of an annular gap induced by an embedded
proto-planet on disk scattered light images and the infrared
spectral energy distribution (SED).  We find that the outer edge
of a gap is brighter in the scattered light images than a similar
location in a gap-free disk. The stellar radiation that would have
been scattered by material within in the gap is instead scattered
by the disk wall at the outer edge of the gap, producing a bright
ring surrounding the dark gap in the images.  Given sufficient
resolution, such gaps can be detected by the presence of this
bright ring in scattered light images.  For planets within 10 AU
from the central star, the bright ring will be detectable only by
next generation instruments, so we also briefly discuss the case
of a planet farther out in the disk and its potential detection by
existing instruments.

A gap in a disk also changes the shape of the SED. Radiation that
would have been absorbed by material in the gap is instead
reprocessed by the outer gap wall.  This leads to a decrease in
the SED at wavelengths corresponding to the temperature at the
radius of the missing gap material and a corresponding flux
increase at longer wavelengths corresponding to the temperature of
the outer wall.  We note that, unlike an inner hole in the disk,
the presence of an annular gap does not change the bolometric IR
flux; it simply redistributes the radiation, previously produced
by material within the gap, to longer wavelengths.  This implies
that the changes in the SED generally will be smaller for gaps
than holes. Although it will be difficult on the basis of the SED
alone to distinguish between the presence of a gap and other
physical effects, the level of changes can be sufficiently large
to be measurable with current instruments (e.g., Spitzer).

\end{abstract}


\section{Introduction}

Inner holes in protostellar and planetary debris disks surrounding young 
stars have been detected through the study of their spectral energy
distributions \citep*{koerner, jura98, calvet,R03} and confirmed, in some
cases, by direct images \citep{koerner98, jay, weinberger, schneider99}.
Additionally, warm optically thin gas in the cleared region may be
responsible for the observed CO line emission in T Tauri stars \citep*{najita}.

The presence of such holes has been suggested to be evidence that planets
have formed in these systems \citep*{strom93}.  When planets form in the
protostellar disk, they initially sweep up the material in their vicinity,
tidally clearing gaps in the disk \citep{lin86}.  Ultimately as accretion
stops, the disk dissipates creating a large central hole (see
\citealt*{hollenbach00} for a discussion of possible mechanisms).  Since
gaps and holes in protostellar disks arise through the gravitational
interaction between the disk and planets, detailed studies of the emission
from disks containing planets may allow us to place constraints on the
properties of the embedded planets and on the evolution of these
systems.

Most previous work has investigated the effects of holes rather
than gaps and assumed that the hole or gap simply causes a
reduction in the flux at a particular wavelength due to the lack
of dust at the equilibrium temperatures within the hole
\citep[e.g.,][]{hillenbrand} or gap \citep{boss,beckwith}.  However,
the gap is vertically extended, so one must also account for the 
walls of the gap.  Initial studies that included some of these 
3-D radiative transfer effects were performed by \citet{wolf02,wolf05} 
who combined the results of 2-D hydrodynamics simulations with a 3-D Monte Carlo radiative transfer code to explore the observational consequences of 
a planetary gap.  Since \citet{wolf02,wolf05} were primarily interested in 
detecting the disk density structures induced by the planet, they assumed 
the disk scale height, $H/r$, is independent of radius (i.e., an 
unflared disk) when converting the 2-D surface density from the hydrodynamcs 
simulation into the vertically extended (3-D) density required by the 
radiative transfer model. However, as explored by \citet*{dullemond} and 
\citet{calvet}, the inner wall of a vertically extended disk with a hole is 
directly illuminated by the central star, which increases the temperature
of the wall relative to the upper surface of the disk itself, producing
a bright inner edge to the disk.  A similar effect occurs when a gap is 
created in a {\it flared} disk.

Here we investigate the properties of illuminated gap edges and
how they affect the infrared spectral energy distribution (SED)
and morphology of scattered light images.  In a flared disk, the
outer wall of the gap intercepts and scatters radiation that
normally would have been reprocessed within the gap.  Thus the
outer wall produces a bright ring surrounding the dark gap in the
scattered light images --- an effect not observed by \citet{wolf02} 
because they did not employ a flared disk.  Similarly, the increased 
temperature of the outer wall increases the SED at intermediate 
wavelengths.  To obtain SEDs and scattered light images, we coupled 
the output of $2$-D hydrodynamics simulations of a disk/planet interaction 
\citep*[see, for example,][]{masset03,Varniere04} to our 3-D Monte Carlo radiative transfer (MCRT) codes, assuming a flared disk 
geometry.\footnote{We also have performed full hydrostatic equilibrium 
calculations of the disk vertical structure, but the results are essentially the same as the power law flared disk models.}  The SEDs were calculated using 
the radiative equilibrium method developed by \citet{BW}, as extended
to T Tauri disks \citep[2002b]{wood} and protostellar envelopes
\citep{whitney}.  The scattered light images were calculated using
the MCRT code developed by \citet{whitney92} for young stellar
object (YSO) disks.  Subsequently, this scattered light code was
modified to efficiently calculate images of complex 3-D YSO
envelope geometries \citep{wood01,whitney02}.

\section{Methods}

Our previous work includes a numerical hydrodynamics study of the
properties of disk edges that are maintained by giant planets
\citep{Varniere04}.  The 2-D hydro simulations use a polytropic
equation of state with $\gamma$ $\approx 1$ to mimic the effects 
of strong cooling (reducing $\gamma$ increases the internal degrees 
of freedom thereby increasing the compression behind shocks as would 
occur with strong cooling). In this work we use the results
of two hydro simulations, consisting of 2-D gas surface density
distributions, to compute synthetic scattered light images and
SEDs using the MCRT codes.

Both simulations have the same initial conditions: a central star of 
mass $M_* = 1 M_\odot$, radius $R_\star = 2.5 R_\odot$, effective 
temperature of $4.\ 10^3$K, and
a gas disk extending from $0.1 ~AU < r < 10 ~AU$ with a Reynolds
number (${\cal R} = 10^7$). We use an initial disk surface density
profile $\Sigma(r) \propto 1/r$ and specify the normalization so
that the total disk mass $M_d = 10^{-2} M_\odot$. The first
simulation has no planet and acts as a control. The second
simulation contains a single two-Jupiter-mass-planet ($M_p = 2
M_J$) in a circular orbit of radius $1$ AU. 
We ran the simulations for 1000
orbits of the planet --- long enough that the gap density attains its
steady-state value.

We note that our simulated disks only extend to $10$AU. To predict
longer wavelength emission, we require a larger disk extent, so
after the hydro simulations are performed, we artificially extend
the gas surface density out to a disk radius of $300 ~AU$ using
our $(1/r)$ initial density profile.  Similarly, we extend the surface
density to the disk inner radius $r_i$.  The resulting surface
density then is entered into the MCRT code.

The Monte Carlo radiation transfer code is fully 3-D, so we require an input
3-D disk density.  To convert the 2-D disk surface density (from the hydro
simulation), we assume the vertical structure of the disk is Gaussian with a
scale height that varies as a power law with radius \citep[i.e., a flared
disk; see][section 2.1, for details]{whitney}.  In the future, we hope that
fully 3-D simulations of the disk will allow a more accurate treatment of
the vertical disk structure.  Finally for the dust opacity, we use the large
dust grains that \citet{wood02} found were required to fit the SED of HH30 IRS.

\subsection{Scattered Light Images}

\begin{figure*}
\epsscale{0.60} \plotone{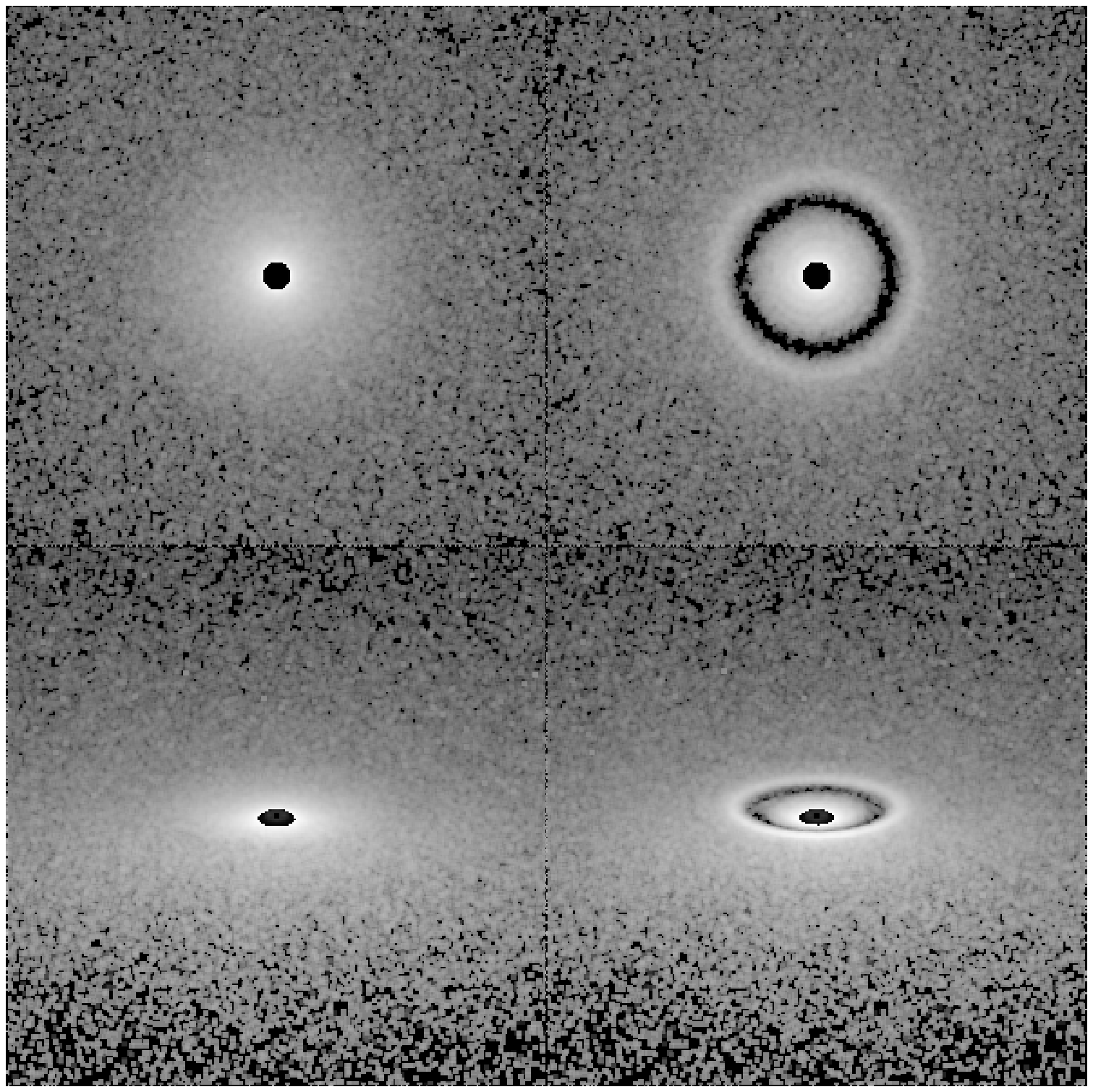} \caption[]{Model scattered light
images. The top figures show the log of the optical disk surface
brightness viewed at an inclination of $i = 5^\circ$. The bottom
figures show disks viewed at $i=70^\circ$. On the left there is no
gap in the disk, on the right there is a gap created by a $2$
Jupiter mass planet clearing a gap centered at 1 AU with a width
of about $1$AU. (Gap width defined via half-minimum depth of
density in gap; \cite{Varniere04}) The images show a region $7
\times 7$ AU. } \label{fig:scat_incli}
\end{figure*}

In Figure~\ref{fig:scat_incli}, we present the scattered light
images produced from the Monte Carlo simulations.  Each image was
created using $10^8$ photons.  Although the photons
were propagated throughout the entire disk (out to 300AU),
Figure~\ref{fig:scat_incli} shows only the central $7 \times 7$
AU. The images at left show the simulations without a planet,
while the images at right include a $2 M_J$ planet located at
$1$AU. The width of the gap (where the density is significantly
lowered) is about $1$ AU. From top
to bottom are shown two inclinations: an almost face-on system
($5^\circ$) and a near edge-on ($70^\circ$) system. As expected,
the gap created by the planet also creates a gap in the scattered
light images.  Furthermore, the outer edge of the gap appears
brighter compared to the same location in the control case.

\begin{figure*}
\epsscale{0.70}
\plotone{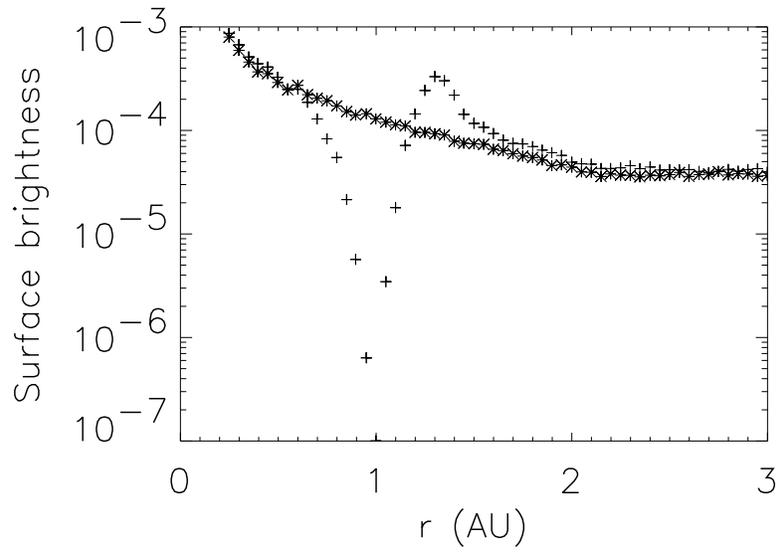}
\caption[]{Comparison between the azimuthally averaged
optical surface brightness profiles for
two disks viewed face-on.
One disk contains a 2 Jupiter mass planet at 1AU (shown as crosses) and the
other lacks a planet (shown as starred points).
The simulated disks are the same as those shown in Fig 1.
The gap is seen as a decrease in the surface brightness profile near the planet.
There is also a bump in the profile on the outer gap edge, corresponding to
the bright ring seen in Fig.~1.
}
\label{fig:lum_comp}
\end{figure*}

To investigate the morphology and magnitude of the excess
emission, we plot in figure \ref{fig:lum_comp} the azimuthally
averaged surface brightness profile for a simulation with a planet
at 1AU viewed face-on. This shows that the ring located at the
outer gap edge has a surface brightness that is about four times
as bright as the emission from the smooth disk at the same
location. This excess emission is likely caused by the increased
illumination of the outer wall of the gap by stellar photons that
are no longer absorbed by disk material within the gap (recall
that the disk scale height increases with radius, so the photons
that would have been absorbed in the gap now hit the outer
vertical wall of the gap).  We conclude that a bright ring
surrounding a darker ring is a characteristic signature of a gap
in a flared disk.  This is in distinction to the solitary dark 
annulus that occurs when disk flaring is not included \citep[see fig.~2 of][]{wolf02}.  Consequently, we expect that the brightness of
the ring might be used to measure the degree of disk flaring.

After the initial gap is created, the inner region of the disk eventually
is cleared, producing an inner hole in the disk.  \citet{wolf05} explored
the transition between these phases (their model has both a gap and
a partially cleared inner disk).  Interestingly, despite the absence
of flaring, they now find a bright ring surrounding a darker gap \citep[see 
fig.~2 of][]{wolf05}.  Most likely, this is because the inner region of 
their disk is optically thin, so the outer wall of the gap is now 
becoming the inner wall of the remaining disk (and inner walls directly 
illuminated by the star also produce bright rings).  We conclude that a bright 
ring surrounding a dark gap will first occur (in scattered light images) 
as the protoplanet initially clears a gap.  Then as the inner disk
is cleared, the wall will become hotter as it begins to be directly
illuminated by the star.  Thus (in thermal emission images) the ring 
will become brighter as the inner hole develops.

We will now focus on the detectability of the gap itself.
The size of the gap opened by a planet depends on the mass of the planet
and its semi-major axis \citep{Varniere04}.
For a $2 M_J$ planet at $1$AU, the gap size is about $1$\ AU.
Taking a typical gap of $1$AU, for a  system at 100 pc ($1\arcsec = 100$AU), 
an angular resolution of
0.2\arcsec\ would allow
a gap to be resolved for a Jupiter planet with $a \gtrsim 25$AU. An angular
resolution of 0.02\arcsec\ would allow a gap to be resolved for planets
with $a \gtrsim 2.5$AU. We conclude that the direct detection of
planetary gaps in the inner regions of the disk will require the
higher resolution of future instruments.  Finally we note that radial
velocity and transit searches, currently the principle means of planet
detection, are biased against the detection of planets
at large distances, making the direct imaging searches we propose here complimentary.

\subsection{SED}

Since current instruments can only detect gaps in the outer
regions of the disk, we are also interested in the potential for
an indirect detection of a planetary gap via its impact on the
SED.  Note, however, that SEDs generally are not unique since
different physical effects can sometimes cause similar changes.
This means that careful modeling of all available observations is
required to limit the range of possibilities before one may
conclude that a gap is the most likely origin of the changes in
the SED.  To study the effect of gaps on the SED, we have performed
two additional sets of simulations.  The initial conditions for 
the first set of simulations places the planet close to the star
at $r_p = 0.15 ~AU$ with the inner edge of the disk located at 
the dust destruction radius, $r_i=0.07~AU$. 
The second set has a planet at $r_p = 5 ~AU$ with
the inner edge of the disk located at $r_i=1~AU$.  In
this later case, it is assumed that a second planet at $r<r_i$ has
cleared out an inner hole.  These two cases bracket the SED
behavior we have observed from a larger set of runs covering a large
range of parameters.  For each
set of simulations we calculated SEDs using two different dust models
that comprise limiting cases: large dust grains used to
succesfully model HH30 \citep{wood02} and ISM dust \citep*{kmh}. 
Both models use a power law distribution of grain sizes from small to large 
(with an exponential cutoff).  The primary difference is that the HH30 dust 
has a higher proportion of big grains.

\begin{figure*}
\epsscale{1.}
\plotone{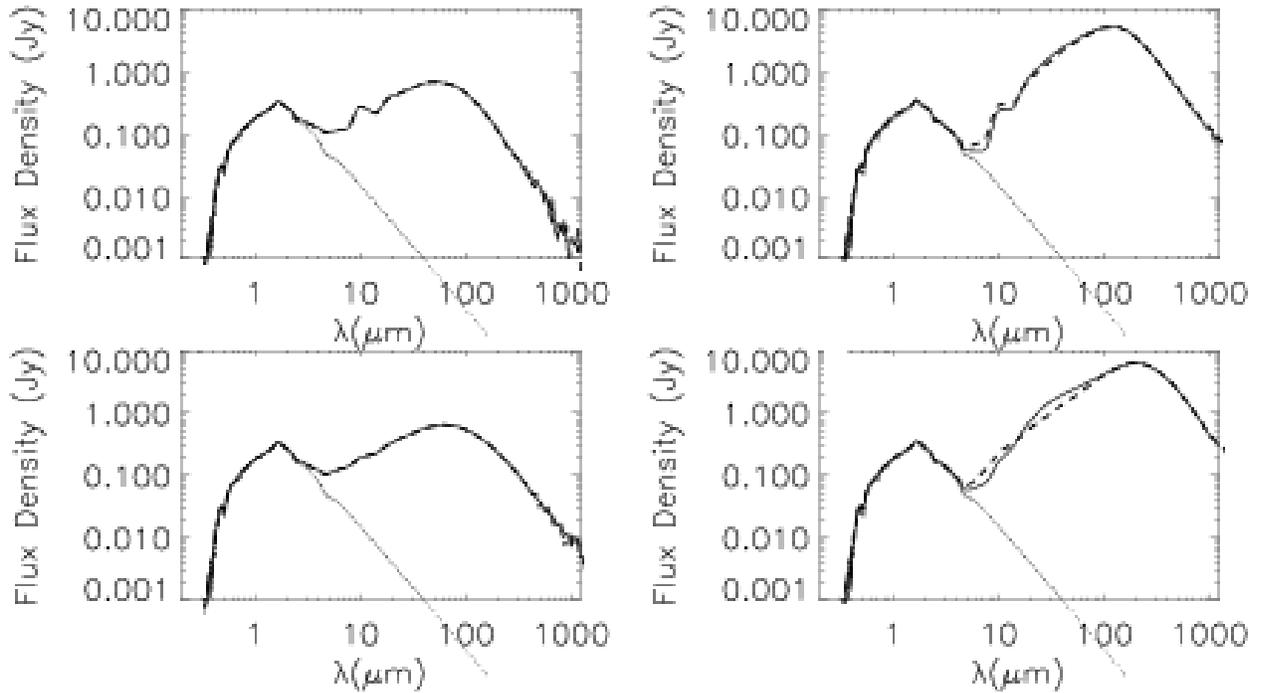}
\caption[]{Spectral Energy Distributions.  The spectral energy distributions
for simulations containing a gap  created by a $2 M_J$ planet are shown
(solid lines) in comparison to a control simulation lacking a planet (dotted lines).  The left panels show the $r_p=0.15~AU,~r_i=0.07~AU$ models, while
the right panels show the $r_p=5~AU,~r_i=1~AU$ models.  In the top (bottom) 
panels, we show the results for the ISM (HH30) dust model.
\label{fig:sed_5}}
\end{figure*}

To calculate the the infrared SED for each of our simulations,
including the control simulations without a gap, we used the Monte
Carlo radiative equilibrium code described by \citet{whitney}. The
SEDs for four cases are displayed in Figure~\ref{fig:sed_5}. The
leftmost panels show the small hole models ($r_i=0.07~AU,~r_p=0.15~AU$);
the rightmost panels show the large hole models ($r_i=1~AU,~r_p=5~AU$).
Similarly, the upper panels employ ISM dust, while the lower panels
use the large HH30 dust grains.  Each plot compares the SED with
a gap (solid line) to that of the control simulation without a gap 
(dashed line).

We focus first on the large hole ($r_i=1~AU$) model with large dust 
grains (lower right panel).  Our results indicate that the presence 
of the gap reduces the emission between $\sim 5\ \mu{\rm m}$ and 
$\sim 20\ \mu{\rm m}$ as expected owing to the removal of the 
emission by warm dust in the gap. 
Around $8\ \mu {\rm m}$ the deficit is about $30$\% of the control
simulation. Note, however, the presence of excess emission at
longer wavelengths (from $\sim 20\ \mu{\rm m}$ to $\sim 100\
\mu{\rm m}$). At $25\ \mu {\rm m}$ this excess is of the order of
$7$\% of the control simulation. This excess arises from the
heating of the vertical disk wall at the outer edge of the gap.
The heating occurs because the photons that would have been
absorbed in the gap are reprocessed instead by the outer wall.
Since this effect does not change the total luminosity reprocessed
by the disk, the gap does not change the IR bolometric flux.
Instead the gap causes the deficit to be redistributed to longer
wavelengths, preserving the total area of the IR SED.  One
consequence of this redistribution is that gaps produce smaller
changes in the SED than previous work suggested (compare
Fig.~\ref{fig:sed_5} to Fig.~3 of \citealt{boss} and Fig.~4 of
\citealt{beckwith}).

The conservation of IR bolometric flux is a primary factor
that can distinguish gaps from inner disk holes. In the case of a small
inner hole, changing the hole radius changes how much stellar radiation 
escapes through the hole.  But for a large inner hole (one for which 
the star is effectively a point source), the situation is less clear; it
depends on how the increase of disk radius, which removes material from 
the disk, alters the solid angle subtended by the optically thick region of the disk.  In general, one expects that increases in hole radius will decrease
the fraction of starlight reprocessed by the disk. Thus holes with different radii will most likely have different IR bolometric 
fluxes.  This implies that increasing the hole radius produces a flux deficit at short wavelengths without producing a correspondingly large flux increase at long wavelengths. In contrast, gaps of different radii (and widths) have the same IR bolometric flux, which also implies that gaps generally produce smaller changes in the SED than holes.

Although, in general, a gap redistributes the flux, the amount of
redistribution (and hence detectability of the gap) depends sensitively 
on both the location of the gap and the hole radius (inner edge of the 
disk). For example, if the disk extends to small radii (left panels
of Fig.~\ref{fig:sed_5}), we 
find that emission from hot dust in the inner regions of the disk 
overwhelms and fills in the potential deficit caused by the gap. 
Similarly, the emission excess becomes too weak to be detected.

Even if the flux changes are large enough to be detected, it will be
very difficult to conclude that a gap is responsible.  For example,
changes in the dust properties will alter the wavelength-dependence of
the opacity, which can affect the shape of the SED.  Comparing the
top and bottom panels on the right side of Figure~\ref{fig:sed_5},
we see that changing the dust size distribution produces changes
in the mid-IR flux levels that are comparable in magnitude to those 
produced by a gap.  Therefore, depending on the details of the dust 
opacity, it is quite possible that these changes could mimic the 
effects of a gap.  Similarly, changing the disk scale height (by 
dust settling, for example) in combination with changing the hole 
radius could also alter the shape of the SED in ways that might mimic
a gap.  Such well-known degeneracies in the parameters of SED model
fitting imply that it will be quite difficult to infer the presence
of a gap on the basis of the SED alone.

\section{Conclusion}

We have studied the observational consequences of the presence of
a gap in a protoplanetary disk using a Monte-Carlo radiative transfer code.
Our results show that disk gaps appear as more than simply dark annuli in the images.  The direct illumination by stellar photons
of the vertical disk wall at the outer edge of the gap results in a
dramatic brightening, thereby producing a characteristic bright ring
around the dark gap in the scattered light images.  

The detectability of the gap will depend on the inclination of the disk, 
the resolution of the detector, and on the size of the gap.  Note that if 
the disk viscosity is higher than we considered here, the gap will be
narrower and may contain more gas, making it more difficult to
identify in either a scattered light image or an SED.

Our results for the SEDs from disks with gaps show once again that
the direct illumination and therefore heating of the outer gap
wall can have measurable consequences (but only in favorable 
circumstances).  We find both an emission
excess (at longer wavelengths) due to the heating of the outer
wall of the gap, as well as an emission deficit (at shorter
wavelengths) due to the cleared region within the gap. Furthermore
unlike an inner disk hole, the gap does not change the fraction of
the stellar luminosity reprocessed by the disk, so the gap does
not change the bolometric IR flux; it simply redistributes the
flux from shorter to longer wavelengths.  A consequence of this
redistribution is that gaps will be harder to detect than inner
holes. Nonetheless, the changes in the SED can be detected in some
cases, but we caution that other physical effects, such as variations in 
the disk thickness, density distribution, and dust properties may change the SED in ways that mimic the presence of a gap.  We conclude that detecting gaps on the basis of the SED alone in optically thick disks will
be difficult, so the most likely method for unambiguously detecting the
earliest stages of planet formation will be to observe a bright ring
surrounding a dark gap in scattered light images.
\acknowledgments
This work was supported by the National Science Foundation: grants
AST-0307686 (AC, JEB),
AST-9702484 (AF), AST-0098442 (AF), AST-0406823 (AQ)
NASA: grants NAG5-8794 (JEB), NAG5-8428 (AF), NNG04GM12G (AQ,AF)
DOE grant DE-FG02-00ER54600 (AF), and the Laboratory for Laser Energetics.


\end{document}